\documentclass[usenatbib,useAMS]{mn2e}
\usepackage{epsfig,color}
\usepackage{amsmath}
\usepackage{mathptmx}
\usepackage{amssymb}
\usepackage{units}
\usepackage{url}
\usepackage{booktabs}

\newcommand{\apjl}{ApJL}
\newcommand{\apj}{ApJ}
\newcommand{\aap}{A\&A}
\newcommand{\apjs}{ApJS}

\newcommand{\mnras}{MNRAS}

\newcommand{\aj}{AJ}
\newcommand{\nat}{Nature}
\newcommand{\prc}{Phys. Rev. C}
\newcommand{\gcc}{\ \mathrm{g\ cm^{-3}}}

\newcommand{\keV}{\,\mathrm{\lowercase{ke}V}}

\newcommand{\nuc}[2]{\ensuremath{\mathrm{^{#1}#2}}}
\newcommand{\ye}{\ensuremath{Y_\mathrm{e}}}
\newcommand{\msun}{\ensuremath{\mathrm{M}_\odot}}

\title[5.9 keV X-rays in SNe Ia]{5.9 keV Mn K-shell X-ray luminosity from the
decay of $\mathbf{^{55}{Fe}}$ in Type~Ia supernova models}
\author[Seitenzahl et al. 2014]
{\parbox{\textwidth}{I.~R.~Seitenzahl,$^{1,2,3}$\thanks{E-mail:
ivo.seitenzahl@anu.edu.au}
  A.~Summa,$^{1,2}$
  F.~Krau\ss,$^{1,4}$
  S.~A.~Sim,$^{5}$
  R.~Diehl,$^{6}$
  D.~Els\"asser,$^{1}$
  M.~Fink,$^{1}$
  W.~Hillebrandt,$^{2}$ 
  M.~Kromer,$^{2,7}$
  K.~Maeda,$^{8,9}$
  K.~Mannheim,$^{1}$
  R.~Pakmor,$^{10}$
  F.~K.~R\"opke,$^{1}$
  A.~J.~Ruiter,$^{2,3}$
  J.~Wilms$^{4}$}\vspace{0.4cm}\\
\parbox{\textwidth}{$^{1}$Institut f\"ur Theoretische Physik und Astrophysik,
Universit\"at
  W\"urzburg, Emil-Fischer-Stra{\ss}e 31, 97074 W\"urzburg, Germany\\
$^{2}$Max-Planck-Institut f\"ur Astrophysik,
  Karl-Schwarzschild-Stra{\ss}e 1, 85748 Garching, Germany\\
$^{3}$Research School of Astronomy and Astrophysics, Mount Stromlo Observatory, Cotter Road, Weston Creek, ACT 2611, Australia\\
$^{4}$Dr.\ Karl Remeis Observatory \& ECAP, Sternwartstr.\ 7, 96049 Bamberg,
Germany\\
$^{5}$Astrophysics Research Centre, School of Mathematics and Physics, Queen's
University Belfast, Belfast BT7 1NN, UK\\
$^{6}$Max-Planck-Institut f\"ur extraterrestrische Physik,
  Giessenbachstra{\ss}e, 85748 Garching, Germany\\
$^{7}$The Oskar Klein Centre \& Department of Astronomy,
  Stockholm University, AlbaNova, 
  SE-106 91 Stockholm, Sweden\\
$^{8}$ Department of Astronomy, Kyoto University, Kitashirakawa-Oiwake-cho,
Sakyo-ku, Kyoto 606-8502, Japan\\
$^{9}$Kavli Institute for the Physics and Mathematics of the
  Universe (WPI), Todai Institutes for Advanced Study (TODIAS),
University of Tokyo, \\5-1-5 Kashiwanoha, Kashiwa,
  Chiba 277-8583, Japan\\
$^{10}$Heidelberger Institut f\"{u}r Theoretische Studien,
  Schloss-Wolfsbrunnenweg 35, 69118 Heidelberg, Germany}
}

\date{\today}
\pubyear{}
\begin{document}
\maketitle
\begin{abstract}
We show that the X-ray line flux of the Mn K$_\alpha$ line at
  $5.9\,\mathrm{keV}$ from the decay of \nuc{55}{Fe} is a promising
diagnostic to distinguish between Type Ia supernova (SN Ia) explosion
models. Using radiation transport calculations, we
compute the line flux for two three-dimensional explosion models:
a near-Chandrasekhar mass delayed
detonation and a violent merger of two (1.1 and 0.9 \msun) white
dwarfs. Both models are based on solar metallicity zero-age
  main sequence progenitors. Due to explosive nuclear burning at higher
density, the delayed-detonation model synthesises ${\sim}3.5$ times more
radioactive \nuc{55}{Fe} than the merger model. 
As a result, we find that the peak Mn K$_\alpha$ line flux of the
delayed-detonation model exceeds that of the merger model by a factor of ${\sim}4.5$. 
Since in both models the $5.9\,\mathrm{keV}$ X-ray flux peaks five to six years
after the explosion, a single measurement of the X-ray line emission
at this time can place a constraint on the explosion physics that
is complementary to those derived from earlier phase optical
spectra or light curves.
We perform detector simulations of current and future X-ray telescopes to investigate the possibilities 
of detecting the X-ray line at $5.9\,\mathrm{keV}$. Of the currently
existing telescopes, \textsl{XMM-Newton}/pn is the best instrument for
close (${\lesssim}1-2\,\mathrm{Mpc}$), non-background limited SNe~Ia because of its large effective area.
Due to its low instrumental background, \textsl{Chandra}/ACIS is
currently the best choice for SNe~Ia at distances above ${\sim}2\,\mathrm{Mpc}$. For the delayed-detonation scenario, a line detection is feasible with
  Chandra up to ${\sim}3\,\mathrm{Mpc}$ for an exposure time of
  $10^6\,\mathrm{s}$. We find that it should be possible with currently existing X-ray
  instruments (with exposure times \mbox{${\lesssim}5 \times
    10^5\,\mathrm{s}$}) to detect both of our models at sufficiently high S/N to
  distinguish between them for hypothetical events within the Local
  Group. The prospects for detection will be better with future missions. For
example, the proposed \textsl{Athena/X-IFU} instrument could detect
our delayed-detonation model out to a distance of
${\sim}5\,\mathrm{Mpc}$. This would make it possible to study future
events occurring during its operational life at
distances comparable to those of the recent supernovae SN~2011fe
(${\sim}6.4\,\mathrm{Mpc}$) and SN~2014J (${\sim}3.5\,\mathrm{Mpc}$).
\end{abstract}


\section{Introduction}
\label{sec:int}
SNe~Ia are important for a variety of astrophysical research fields.
Besides their relevance as distance indicators in cosmology \citep[e.g.][]{schmidt1998a,riess1998a,perlmutter1999a}, they
are an important ingredient in models of star formation and galaxy
dynamics 
\citep[e.g.][]{scannapieco2008a}.
Furthermore, their nucleosynthesis contributes significantly to the chemical
evolution of galaxies \citep[e.g.][]{burbidge1957a}
and they are possible sources of galactic positrons 
\citep[e.g.][]{clayton1973a} and gamma rays \citep[e.g.][]{clayton1969a}.
Despite their significance, a clear picture of the progenitor systems and
explosion mechanisms responsible for SNe~Ia continues to elude us.
It is generally agreed that 
thermonuclear explosions of carbon--oxygen white dwarfs (WDs) are the origin of SN Ia
explosions, but a variety of evolutionary channels leading to such an explosion have
been suggested. For a recent review of explosion scenarios and explosion models see e.g.\ \citet{hillebrandt2013a}. 

Although SNe~Ia are now routinely observed in the optical part of the
electromagnetic spectrum, a clear distinction between competing explosion
scenarios based only on the optical emission of the SNe is difficult
\citep[e.g.][]{roepke2012a}.  Recently, there has been a revived effort
 to identify additional signatures, such as the contribution to the early-time
light curve from a possible companion
\citep[e.g.][]{kasen2010a, bloom2012a, brown2012a}, Na I D line absorption features
interpreted as evidence for circumstellar material
\citep{patat2007a,sternberg2011a,dilday2012a}, gamma-ray line and continuum emission
\citep[e.g.][]{sim2008a, maeda2012b, summa2013a}, or 
the chemical evolution of Mn in the Galaxy \citep{seitenzahl2013b}, 
to distinguish between different explosion models.

X-ray emission from SNe~Ia is another, although still unobserved, signature of the physical processes occurring in these explosions.
X-rays in SNe~Ia are produced in several ways, including bremsstrahlung of fast recoil Compton electrons, down-scattering of gamma rays, as well as the relaxation of ionised atoms following photoelectric absorption, Compton scattering, collisional ionisation or electron capture decay \citep[e.g.][]{burrows1990a, clayton1991a, the1994a}. 
\citet{the2014a} show that SN~2014J in M82 is sufficiently close that \textsl{NuSTAR} \citep{harrison2013a}, \textsl{\mbox{ASTRO-H}} \citep{kokubun2010a}, and perhaps \textsl{INTEGRAL} \citep{lebrun2003a} could detect the emerging hard X-ray emission and possibly constrain explosion models with the data. Incidentally, SN 2014J is the first SN~Ia for which gamma rays from the decay of \nuc{56}{Ni} \citep{diehl2014a} and \nuc{56}{Co} \citep{churazov2014a} have been observed (with \textsl{INTEGRAL}).
However, although \citet{the2014a} extensively analyse the prospects of detecting X-rays from SN~2014J, they do 
not discuss the Mn K$_\alpha$ line from the decay of \nuc{55}{Fe}.
\citet{leising2001a} already pointed out that X-ray signals of SN electron
capture radionuclides such as \nuc{55}{Fe} can be used as a diagnostic of SN nucleosynthesis
by directly measuring isotopic abundances. A search for the \nuc{55}{Mn} K$_\alpha$ line in $400\,\mathrm{ks}$ of \textsl{Chandra}/ACIS data
of SN~1987A only produced upper limits \citep{leising2006a}. 
However, \citet{leising2001a} already noted that SNe~Ia (from exploding near-Chandrasekhar mass WDs) 
synthesise significantly more \nuc{55}{Co} than core-collapse SNe, which enhances prospects of detection.

In this paper, we revisit the detection prospects of the X-ray emission directly related to the radioactive decay of \nuc{55}{Fe} produced in SNe~Ia.
These X-rays are emitted when the atomic electron configuration of the daughter nucleus
\nuc{55}{Mn} relaxes to eliminate K-shell vacancies.
Electron capture of \nuc{55}{Fe} produces a line
doublet with energies of $5.888\,\mathrm{keV}$ ($8.2\,\%$) and
$5.899\,\mathrm{keV}$ ($16.2\,\%$) \citep{junde2008a}. Using three-dimensional
hydrodynamical simulations and subsequent radiative transfer calculations, we
determine the respective line fluxes for two explosion models -- a violent merger of two WDs
with a sub-Chandrasekhar mass primary, and a delayed detonation in a near-Chandrasekhar mass WD.
For the latter, we use the N100 model from \citet{seitenzahl2013a}. 
For the violent merger of two WDs
(1.1 and 0.9\,\msun), we use the model published in \citet{pakmor2012a}.
These two explosion models have been shown to reproduce 
many of the observable characteristics of ``normal'' SNe~Ia \citep{roepke2012a,sim2013a}
and are representatives of the single degenerate and double degenerate evolutionary
channels, respectively.
Details of the simulation techniques can be found in the original
publications and references therein. We have chosen these two
explosion models as illustrative examples that highlight the
dependence of the \nuc{55}{Fe} yield on the explosion scenario and primary WD mass. 
Both models employ a fully three-dimensional treatment of the explosion hydrodynamics, 
which, compared to lower dimensional models, establishes a more realistic description of the 
distribution of the radioactive isotopes in the ejecta and enables radiative transfer calculations for different lines of sight.
We consider the use of three-dimensional models important for this problem for the following reason: driven by buoyancy forces, hot and less dense ashes of the nuclear burning ``float towards the surface", that is, they travel in mass coordinate against the gravitational field lines, their places filled by unburned material from down-drafts. This physical behaviour is present in our 3D model but is completely absent in 1D models, where the symmetry constraint makes convective motion impossible and forces the burning products to an artificial ``onion-like" shell structure (where they remain layered in the same order they were burned). This leads to differences in the location of important isotopes: for example, comparing the mass distribution of the N100 \citep{seitenzahl2013a,seitenzahl2014a} and say W7 \citep{iwamoto1999a} models, we see that the Mn (coming from \nuc{55}{Co}), predominantly arising from "normal" freeze-out from nuclear statistical equilibrium (NSE), sits in qualitatively different positions, that is, centrally concentrated at low velocity in the W7 and roughly speaking in a spherical shell at intermediate velocities in our 3D models.
We then use the results of the radiative transfer calculations as input for X-ray telescope signal simulations and investigate the prospects
for distinguishing the explosion models for several current and proposed X-ray
astronomy missions.

For explosive nucleosynthesis conditions occurring in SNe~Ia, 
only a small fraction of the \nuc{55}{Fe} ($t_{1/2}=2.7\,\mathrm{yr}$)
that is present in the ejecta a few weeks after the explosion was 
produced as ``primary'' \nuc{55}{Fe}.  
Most of the \nuc{55}{Fe} present at late times was synthesised as 
\nuc{55}{Co} \citep[e.g.][]{truran1967a}.
\nuc{55}{Co} decays with a half-life of $17.5\,\mathrm{h}$ to \nuc{55}{Fe}
and is mainly produced for electron fractions $\ye \lesssim 0.5$ in two distinct
processes: ``normal'' freeze-out from NSE and
incomplete Si-burning. For freeze-out from NSE to be ``normal'', the mass
fraction of $\nuc{4}{He}$ during the freeze-out phase has to remain rather low
\citep[$\lesssim$1 per cent, e.g.][]{woosley1973a}. This is the case for
explosive nuclear burning at
relatively low entropy, which implies high density \citep[$\rho \gtrsim 3 \times
10^8 \gcc$, cf.][]{thielemann1986a, bravo2012a}. 

In our delayed detonation of a near-Chandrasekhar mass
WD, densities sufficiently high to undergo ``normal'' freeze-out from
NSE are realised. In contrast, the merger
model has much lower peak densities and predominantly
synthesises \nuc{55}{Co} via incomplete Si-burning \citep[the \nuc{55}{Co}
present in NSE for such low central density models is mostly destroyed 
during the alpha-rich freeze-out via 
$\nuc{55}{Co}(p,\gamma)\nuc{56}{Ni}$, cf.][]{jordan2003a}. 
A recent study has shown that the \nuc{55}{Co} to \nuc{56}{Ni}
production ratio is rather insensitive to nuclear reaction rate
uncertainties \citep{parikh2013a}. Thus, our models should make a
rather robust prediction of different \nuc{55}{Co} 
yields for the two explosion scenarios 
(at equal \nuc{56}{Ni} masses, the abundance of \nuc{55}{Co}
is significantly higher for the delayed detonation), which also drives the
different predicted behaviour of late time bolometric light curves
\citep[][]{seitenzahl2009d,seitenzahl2011b,roepke2012a} and the 
different Mn production \citep{seitenzahl2013b} of these two explosion models. 

\section{Radiative Transfer Simulations}
\label{sec:rt}
We compute the photon flux (photons\,cm$^{-2}$\,s$^{-1}$\,keV$^{-1}$) in the $5.9\keV$ Mn
K$_{\alpha}$ line at energy $\epsilon$
at time $t$ (relative to explosion) for an observer orientation specified by
unit vector
$\mathbf{n}$ using
\begin{equation}
\begin{split}F(t,\mathbf{n},\epsilon) = \ & \frac{p_{\gamma}\ln 2}{4 \pi D^2 \epsilon_0 m_{55} t_{1/2}}
\exp\left(-\frac{t \ln 2}{t_{1/2}}\right) \\[3mm]
& \times \int_{V} 
\rho(\mathbf{r},t)
\,\delta \left( \frac{\epsilon}{\epsilon_0}-1- \frac{\mathbf{n}\cdot\mathbf{r}}{ct} \right) 
\,X_{55}^0(\mathbf{r},t)  
\,e^{-\tau(\mathbf{r},t,\mathbf{n})} \, \mbox{d}V,
\end{split}
\end{equation}
where the integral runs over the entire volume of the
ejecta and $D$ is the distance of the SN. Here, $t_{1/2} =
2.7$~yr is the half-life of \nuc{55}{Fe} (the
parent of \nuc{55}{Mn}), $p_{\gamma} = 0.244$ \citep{junde2008a} 
is the probability per \nuc{55}{Fe} decay of producing a 
$5.9\keV$ Mn K$_{\alpha}$ photon, $\epsilon_0$ is the rest energy of the line,
$\rho$ is the mass density of the ejecta at position
$\mathbf{r}$, $\delta$ is the Dirac delta function, $m_{55}$ is the atomic mass of \nuc{55}{Co}
and $X_{55}^0$ is the sum of the mass fractions of \nuc{55}{Co} and
\nuc{55}{Fe} at $t=0$ (i.e.\ immediately after explosion). Note
that we are treating the  \nuc{55}{Co} decay to \nuc{55}{Fe}  as
effectively instantaneous (since, as noted above, its half life is
orders of magnitude shorter).
The optical depth is given by
\begin{equation}
\tau (\mathbf{r},t,\mathbf{n}) = \sum_{Z=1}^{30} \sigma_{Z}
\int_{\mathbf{r}}^{\infty}
\rho(\mathbf{r}^{\prime},t)\,X_{Z}(\mathbf{r}^{\prime},t) \, \mbox{d}s
\end{equation}
where $\sigma_{Z}$ is the photo-absorption cross-section (cm$^2$~g$^{-1}$) at
$5.9\keV$ (we use the results from \citealt{henke1993a}),
and $X_{Z}$ the mass fraction of the element with atomic number $Z$\@.
The line integral runs along the ray defined by
$\mbox{d}\mathbf{r}^{\prime} =\mathbf{r} + \mathbf{n} \; \mbox{d}s$
from the starting point $\mathbf{r}$ to the outer edge of
the ejecta. Our approach ignores light-travel time effects and assumes that
photo-absorption is the dominant opacity -- i.e.\ electron scattering is
relatively unimportant, which is a good approximation
in SNe~Ia ejecta at the epochs of interest (several years post explosion).

To evaluate equations (1) and (2), we use the ejecta properties 
provided by our hydrodynamic explosion
models. Specifically, as in e.g.\ \cite{kromer2010a}, we use the distributions of
density and
composition reconstructed (on uniform Cartesian grids) via a
smoothed-particle-hydrodynamics-like
algorithm 
from the ensemble of nucleosynthesis tracer
particles at the final state of the explosion simulations ($t =
100\,\mathrm{s}$). 
For all later times it is assumed that the ejecta follow a
homologous expansion law.

\begin{figure}
  \vspace{-0.3cm}
  \centering
  \includegraphics[width=0.9\columnwidth]{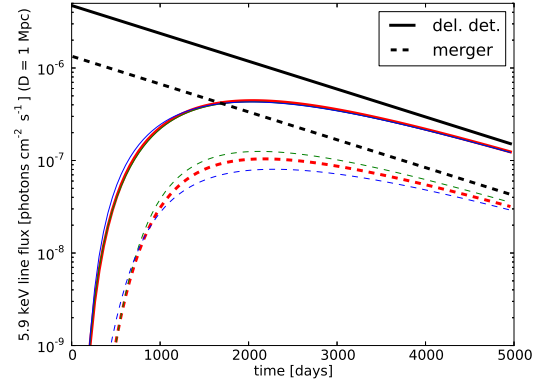}
  \caption{X-ray line flux at $5.9\keV$ for the two different explosion
    models at a distance of $1\,\mathrm{Mpc}$. The solid (dashed) lines show the
    delayed-detonation (merger) model. The three different colours indicate three orthogonal lines of sight.
    Black lines indicate the optically thin limit (no absorption).
}
  \label{fig:f1}
\end{figure}

The results of our radiative transport calculations for the two different
explosion scenarios are given in Fig.~\ref{fig:f1}, which shows the total line flux obtained by integrating equation 1 over photon energy, $\epsilon$.
The $5.9\,\mathrm{keV}$ line flux is significantly larger for the 
delayed detonation than for the violent merger due to the greater mass
of synthesised \nuc{55}{Co}. 
The time evolution of the X-ray line
flux is similar in both cases. Due to the relatively long half-life of
\nuc{55}{Fe} and the large photoelectric opacity of the ejecta to
X-rays at early times, it takes
roughly $2100\,\mathrm{d}$ to reach the maximum fluxes. This also defines the optimal 
time frame for X-ray observations of the $5.9\,\mathrm{keV}$ emission line. 
The influence of attenuation effects can be clearly seen in comparison to the X-ray ``free-streaming'' limits
(see Fig.~\ref{fig:f1}). Photoelectric opacity continues to be
relevant in both models until ${\sim}5000\,\mathrm{d}$.

For both models, we plot the evolution of the line fluxes for three orthogonal lines of sight
to the explosion. As inferred from the colour-coded set
of curves for each model, the effect of different viewing angles is negligible
for the delayed-detonation and moderate for the violent-merger model. There, the
asymmetric ejecta structure and the inhomogeneous distribution of the radioactive 
isotopes \citep[cf.][]{pakmor2012a} lead to a larger (but still
modest) spread between the flux values. However, unlike in gamma rays \citep[cf.][]{summa2013a},
there is no degeneracy between the two models for different viewing angles.

\section{Observability of the 5.9\,\lowercase{ke}V line}
\label{sec:obs}
\begin{figure}
  \includegraphics[width=0.9\columnwidth]{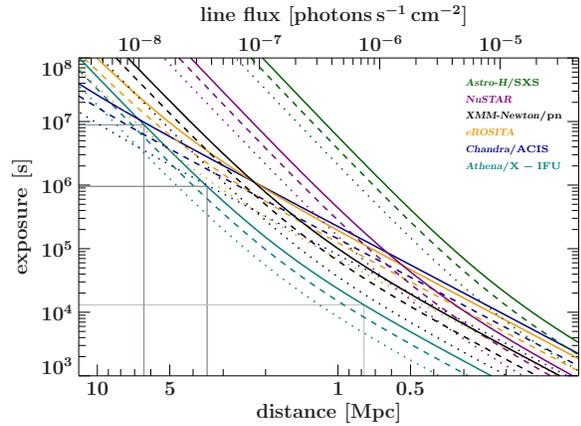}
  \caption{Exposure times required for the detection of the $5.9\keV$ line with
     different X-ray instruments as a function of photon flux (upper
     X-axis) for the delayed-detonation model.  
     The line style (dotted, dashed, solid) encodes the significance
     of the line detection (3, 4, and $5\,\sigma$
     respectively).
    For the maximum X-ray line flux predicted by the model at
    ${\sim}2100$\,d after explosion, 
    the lower X-axis shows the corresponding distances to the SN explosion.
    For a distance of 6.4\,Mpc (distance to SN~2011fe), 3.5\,Mpc (distance to SN~2014J),
     and 0.78\,Mpc (distance to M31), the thin vertical lines mark the required
    exposure times for a $5\,\sigma$ detection of the $5.9\keV$ line with
    respect to the most sensitive instrument in each case.}
  \label{fig:f2}
\end{figure}

In this section, we discuss the detectability of the
$5.9\,\mathrm{keV}$ emission line by current and future X-ray observatories.
In addition to the expected signal from the SN~Ia X-ray line emission, 
we consider the instrumental background and
X-ray continuum contributions.  

There are several potential mechanisms that could produce X-ray
continuum emission at keV energies.
First, there is the bremsstrahlung of fast recoil electrons 
that scatter from ions during their deceleration to
thermal energies within the SN interior \citep{clayton1991a}. These
recoil electrons are produced in Compton-scattering events with the primary
gamma-ray photons of the radioactive isotopes in the ejecta.
The X-ray flux due to this mechanism is most intense in the first few
weeks after the explosion when the radioactive energy deposition of \nuc{56}{Ni} and
\nuc{56}{Co} decay in the expanding SN
is still large and the production of fast electrons near the surface is
efficient because the Compton optical depth remains high.
This changes at later times when the ejecta become fully
  transparent to gamma rays and the production of fast electrons is no
  longer efficient. An extrapolation of the results of \citet[][]{clayton1991a} for the W7 model shows that the internal bremsstrahlung contribution can be completely neglected at the epochs
relevant for this work.

Second, there may be X-ray emission arising from the interaction of the
ejecta with the surrounding circumstellar medium (CSM). Both thermal and non-thermal
emission mechanisms can contribute. Since the X-ray emission of
  shell-type SN remnants usually shows a dominant contribution of
  thermal emission processes \citep[e.g.][]{Ballet2006,badenes2010a},
  we approximate the continuum emission with a thermal bremsstrahlung
  model.
Models assuming typical densities for the ambient medium predict that the thermal X-ray emission from the heated ejecta and ambient material does not reach its
maximum luminosity until several hundred years after explosion
\citep{badenes2003a}. In agreement with such models, observations
generally place only upper limits of the early thermal X-ray emission
of SNe~Ia \citep[e.g.][]{Hughes2007a}.
The intensity of the X-ray emission from CSM interaction depends on
the strengths of the forward and reverse shocks that build up after
the ejecta are launched.
Although initially weak, the X-ray radiation in the shocked regions
increases as more heated and compressed material is accumulated.
To take the effects of CSM interaction into account, we adopt an
analytic model of the underlying X-ray continuum emission 
\citep{Immler2006,Hughes2007a}. 

For the thermal bremsstrahlung, we use $kT =
10\keV$ \citep[cf.][]{Immler2006,Fransson1996}.
The normalisation of the bremsstrahlung emission is given by
the emission measure
\begin{equation}
\mathrm{EM} = \int_{V_{\mathrm{sh}}} n_e n_i~ \mathrm{d}V,
\end{equation} 
where $V_{\mathrm{sh}}$ is the volume of the shocked and emitting material, 
$n_e$ the electron density and $n_i$ the ion density.
Assuming a constant-density ambient medium with
$\rho_\mathrm{AM}=10^{-24}\,\mathrm{g\,cm^{-3}}$, which is a typical number
in most of the Galaxy and the Magellanic Clouds, hydrodynamical simulations
of the shock interactions result in typical emission measures of
$10^{51}\,\mathrm{cm^{-3}}$ to $10^{52}\,\mathrm{cm^{-3}}$ 
at six years after the explosion \citep[cf.][]{badenes2003a}. These assumptions
concerning the ambient medium and the emission measure 
are further justified by the fact that only 
a rare subclass of SNe Ia shows indications for a strong interaction with 
their CSM \citep{hamuy2003a,Russell2012a,dilday2012a,Silverman2013a}.
For this work, we perform all calculations assuming
an emission measure of $10^{51}\,\mathrm{cm^{-3}}$.

We absorb the emission line and the 
thermal bremsstrahlung component by a Galactic neutral hydrogen column \citep[cf.][]{Wilms2000}
of $N_{\mathrm{H}} = 1.8\times10^{20}\,\mathrm{cm}^{-2}$.
This value corresponds to the viewing direction of M101 \citep{Kalberla2005} and is typical for an observation
that does not point towards the Galactic disk. However, at the line energy of 5.9 keV, absorption effects are 
negligible for column densities below ${\sim}10^{22} \mathrm{cm}^{-2}$ and our results therefore do not depend on the exact value of $N_{\mathrm{H}}$.

Next, we estimate the detectability of the 5.9\,keV emission line,
which we assume to be monochromatic for the purpose of this excercise.
The detection significance ${S}/{\sigma_s}$ is given by
  \begin{equation}
    \dfrac{S}{\sigma_s} = r_s \dfrac{\sqrt{\delta t}}{\sqrt{r_s+2r_b}}
  \end{equation}
  with the source count rate $r_s$, the background count rate $r_b$,
  and the exposure time $\delta t$ \citep{Bradt:2004}.
  The background count rate includes instrumental background and
  continuum count rates.

For the detector simulations, we use the effective areas, redistribution matrix functions, and
background count rates of \textsl{Chandra}/ACIS \citep{chandrahb}, \textsl{eROSITA} \citep{erositabg},
\textsl{XMM-Newton}/pn \citep{2003A&A...409..395R}, \textsl{NuSTAR} \citep{harrison2013a}, \textsl{Astro-H}/SXS \citep{astrohb}, and \textsl{Athena}/X-IFU \citep{AthenaX-rayObservatory2014}.
Fig.~\ref{fig:f2} shows the exposure times required by
these six instruments to detect the line at maximum luminosity 
for the delayed-detonation explosion model with 3, 4, and $5\,\sigma$
significance as a function of the distance to the SN\@.
The upper X-axis shows the corresponding photon flux.
The distance is given under the assumption that an observation is
taken at the epoch corresponding to the maximum line flux of the models
(${\sim}2100$\,d after the explosion). 
At this time, the violent-merger model has a line flux that is lower
by a factor of 4.5 compared to the delayed-detonation model. Therefore
the distances corresponding to a detection at a chosen significance
level are shifted by a factor of $1/\sqrt{4.5}$. Equivalently, for
a fixed distance, longer exposure times are required to detect the
line with the same significance as for the delayed-detonation model.

\begin{figure}
    \includegraphics[width=8cm]{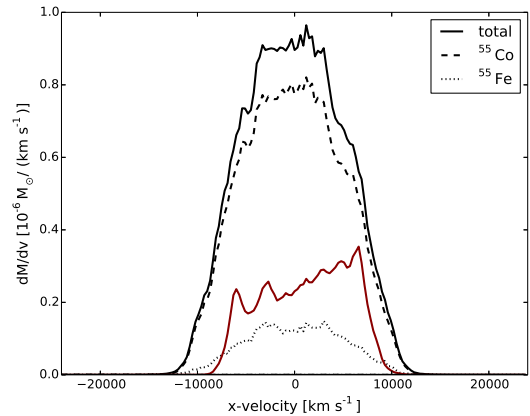}
  \caption{Solid lines are the distribution of the combined mass of \nuc{55}{Co} and \nuc{55}{Fe} in velocity space projected along a line of sight (simulation X-axis); the delayed-detonation model in black and the merger model in red. For the delayed-detonation model, \nuc{55}{Co} (black dashed) and \nuc{55}{Fe} (black dotted) are also shown separately.}
  \label{fig:f3}
\end{figure}

\begin{figure}
    \includegraphics[width=8cm]{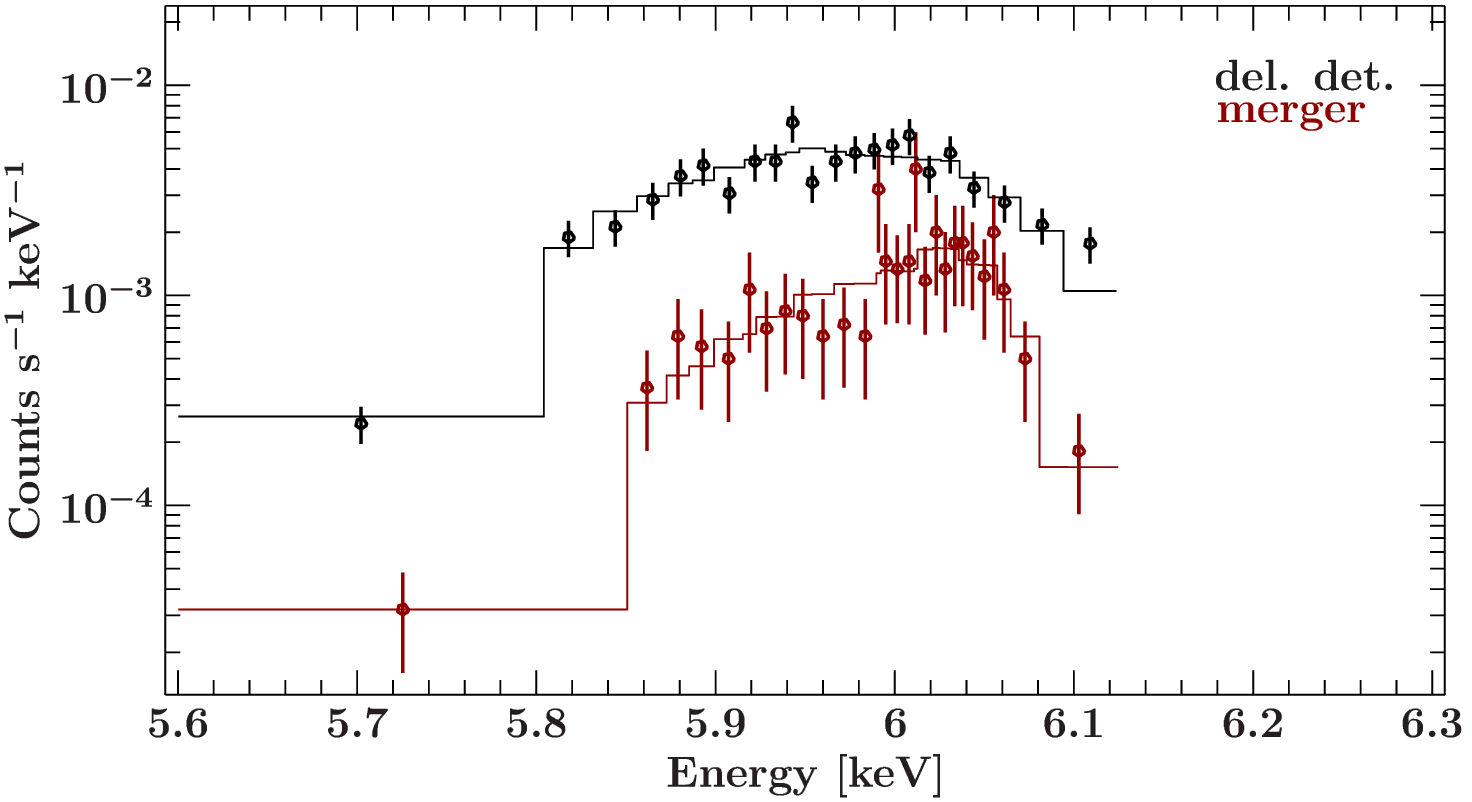}
  \caption{Simulated $500\,\mathrm{ks}$ \textsl{Athena}/X-IFU
      background-subtracted spectra of the 5.9\,keV emission line at
      a distance of 0.78\,Mpc for the delayed-detonation (black) and the violent-merger model (red). 
      The symbols show the simulated data and the solid lines show the best fit of the input model to the data.
      The continuum count rate (emission measure of
      $10^{51}$\,cm$^{-3}$) is below the instrumental background.}
  \label{fig:f4}
\end{figure}

For low line fluxes (and hence large distances), the background count rates in the
detectors dominate the source flux. Due to the low instrumental
background, \textsl{Chandra}/ACIS is the best
choice of the currently operating X-ray telescopes for SNe at distances above $2\,\mathrm{Mpc}$.
For the delayed-detonation model, line detection for close
explosion events in the Local Group up to the Andromeda (M31) distance 
of $0.78\,\mathrm{Mpc}$ is possible for all discussed instruments except \textsl{Astro-H}/SXS, with 
realistic exposure times (${\lesssim} 5 \times 10^5\,\mathrm{s}$).
For source fluxes \mbox{${\gtrsim} 10^{-7}$\,photons\,s$^{-1}$\,cm$^{-2}$},
count rates are no longer background dominated and
\textsl{XMM-Newton}/pn is the best-suited existing instrument because of its large effective area.
For M31 (see Fig.~\ref{fig:f2}), the large effective area
of \textsl{XMM-Newton}/pn leads to exposure times of several 100\,ks only.
The realisation of the currently proposed \textsl{Athena} mission
  would significantly increase the distance limits for a line
  detection. With Athena, the delayed-detonation model could be
  detected (at $3\,\sigma$) up to distances of 5\,Mpc within reasonable exposure times of ${\sim}10^6\,\mathrm{s}$.

Our radiative transfer calculations take the Doppler broadening arising from the velcocity distributions of \nuc{55}{Co} and \nuc{55}{Fe} along the line of sight (see Fig.~\ref{fig:f3}) into account. We can therefore simulate line detection with \textsl{Athena}/X-IFU and \textsl{XMM-Newton}/pn in greater detail, including the broadening and shape for a fully resolved line. An example of the simulations of the proposed \textsl{Athena}/X-IFU
  detector is shown in Fig.~\ref{fig:f4}, where we assume a distance
  of 0.78\,Mpc (distance to the Andromeda Galaxy M31) and an exposure time of 500\,ks. For such an exposure
  time, the delayed-detonation model could be detected  
 up to a distance of $\sim$2.9\,Mpc and the merger model up to a distance of $\sim$1.4\,Mpc by \textsl{Athena}/X-IFU at the $5\,\sigma$ level
 and a clear distinction of the two models is possible. With
  \textsl{XMM-Newton}/pn, our models can both be distinguished for distances
  below $\sim$1.8\,Mpc in case of a 500\,ks exposure.

For the distance of
  $3.5\,\mathrm{Mpc}$ to M82, which is the host galaxy of the recently discovered SN~2014J
  \citep{fossey2014a}, line detection for our models is within reach. 
  For our delayed-detonation 
  scenario (see Fig.~\ref{fig:f2}), we estimate a better than
  $3\,\sigma$ detection with \textsl{Chandra}/ACIS for exposure times of
  $10^6\,\mathrm{s}$. $6.4\,\mathrm{Mpc}$ corresponds to the distance of the recently
  discovered SN~2011fe. A  $5\,\sigma$ line detection for our delayed-detonation
  model with \textsl{Chandra}/ACIS would
  require exposure times of ${\sim}10^7\,\mathrm{s}$, which means that
  SN~2011fe is currently beyond the reach of X-ray telescopes.

The quantitative statements we make about the detectability of SNe~Ia
are of course model dependent. In particular, the assumed metallicity of the zero-age main
sequence (ZAMS) progenitor has a direct effect on the production of
the slightly neutron rich radionuclides \nuc{55}{Co} and \nuc{55}{Fe} (see
Table~\ref{tab:1}). The metallicity dependent yields for the N100 model are taken from \citet{seitenzahl2013a}, who 
approximately take the effect of ZAMS metallicity into account by making the simplyfing assumption that all metals in the ZAMS progenitor are locked up in CNO, 
which is efficiently converted to \nuc{14}{N} during H-burning and then to \nuc{22}{Ne} during core He-burning.
For the merger case, we use the yields from the 1.1 + 0.9 \msun\ model from \citet{pakmor2012a}, which treated ZAMS metallicity in the same way.\footnote{Note that the yields for the three sub-solar metallicity cases of the merger model were determined according to a slightly updated prescription described in \citet{kromer2013b}.}
For the merger model, the effect of the assumed ZAMS metallicity on Mn is quite pronounced, resulting in a strong suppression (by more than an order of magnitude) in the combined mass of \nuc{55}{Co} and \nuc{55}{Fe} when going from solar
to 0.01 times solar metallicity (note that SNe~Ia from progenitors with such low ZAMS metallicity should
be rare in the local Universe based on their predicted age from
theoretical delay-time distributions). In contrast, the
delayed-detonation model based on the explosion of a 
near-Chandrasekhar mass WD exhibits only a modest reduction of less than a
factor of two with decreasing ZAMS progenitor metallicity from solar
to 0.01 times solar. This qualitatively different dependence on
progenitor metallicity is explained by the different nucleosynthetic
processes operating. The high-density regions of the 
near-Chandrasekhar mass delayed-detonation model undergo significant 
in-situ neutronisation via electron capture processes, which explains
the relatively flat dependence on pre-explosion neutron enrichment.
On the other hand, the much lower peak densities of the significantly
sub-Chandrasekhar mass WDs in the merger model prohibit significant neutron
enrichment via in-situ electron captures, which explains the
comparatively steep dependence on pre-explosion neutron enrichment and
hence metallicity. But even if the variations in the abundances of \nuc{55}{Co} and \nuc{55}{Fe}
in dependence on the progenitor metallicities as given in Table~\ref{tab:1} are taken into consideration,
the differences between the two models still amount to at least a
factor of two. In other words, the dimmest of our delayed-detonation
models will still be more than a factor of two brighter than the brightest
of our violent-merger models. In this context we point out that the Mn yields are also rather robust across many near-Chandrasekhar mass explosion models 
\citep[e.g.,][]{iwamoto1999a,seitenzahl2013a} -- the variance between the different models is comparable to the variance introduced by metallicity.
Furthermore, we note that in case of a specific nearby SN Ia, explosion models would be adjusted in order to
reproduce the event as accurately as possible. This includes assumptions concerning the 
metallicity of the progenitor system that may be constrained by information about the environment of the
respective host galaxy. ZAMS metallicity therefore cannot be regarded as a completely free parameter.
 
\begin{table}
  \caption{Combined \nuc{55}{Co} and \nuc{55}{Fe} yields in solar
    masses as a function of progenitor ZAMS metallicity.}
  \centering
  \begin{tabular}{|l|c|c|c|c|} \toprule model name & $1.0\,\mathrm{Z_{\odot}}$ &
    $0.5\,\mathrm{Z_{\odot}}$ & $0.1\,\mathrm{Z_{\odot}}$ & $0.01\,\mathrm{Z_{\odot}}$ \\ \midrule
    N100 (del.\ det.)          & 1.34e$-$2 & 1.11e$-$2 & 8.70e$-$3 & 7.84e$-$3 \\
    1.1\_0.9 (merger)     & 3.85e$-$3 & 2.57e$-$4 & 7.93e$-$5 & 9.46e$-$5 \\
    \bottomrule
  \end{tabular}
  \label{tab:1}
\end{table}

Note that the results of our simulations are quite robust against variations
of the parameters EM, $kT$, and $v$. Due to the fact that the thermal continuum
emission at 2100\,d after explosion is orders of magnitude below the line emission from the decay of \nuc{55}{Fe}, even larger variations 
in EM and $kT$ only have marginal influences on the line detectability.
Therefore, even if the emission measure was enhanced due to higher CSM densities or additional non-thermal processes,
the detectability of the line would remain qualitatively unaffected in case of the major part of normal SNe that do not show strong CSM interaction at early times \citep[cf.][]{Silverman2013a}. 
Changes in the Doppler broadening result in slightly larger error bars for the determination of the line flux
in case of higher velocities. Therefore, the 5.9\,keV emission line can 
be regarded as a unique distinctive feature for the two introduced explosion scenarios,
independent of the choice of a specific parameter set. 

\section{Summary}
\label{summary}
We have calculated the $5.9\,\mathrm{keV}$ Mn K$_\alpha$ line emission
for three-dimensional models of SNe~Ia.
As a result of different central densities at the time of freeze-out from NSE,
the more abundant production of \nuc{55}{Co} leads to a 4.5 times larger
$5.9\,\mathrm{keV}$ maximum line flux in the delayed detonation than
in the merger model. Even taking variations due to the dependence of
the yield of \nuc{55}{Co} and \nuc{55}{Fe} on ZAMS metallicity of the
progenitor into account, the $5.9\,\mathrm{keV}$ signal of the delayed-detonation and the merger 
model remain clearly separated. Therefore, we have shown that
the $5.9\,\mathrm{keV}$ X-ray line provides a distinguishing feature 
between the observable signatures of two leading explosion
scenarios of SNe~Ia: a near-Chandrasekhar mass delayed
detonation and a violent merger of two WDs. 
By performing detector simulations of 
several current and future X-ray instruments, we quantified
the prospects for detecting the $5.9\,\mathrm{keV}$ line and find that, 
due to very low background,
\textsl{Chandra}/ACIS is currently the most suitable instrument for SNe at distances greater than ${\sim}$2\,Mpc. 
Of the existing instruments, \textsl{XMM-Newton}/pn is preferable for distances below ${\sim}$2\,Mpc because of 
the larger effective area. For delayed-detonation SNe~Ia at distances
  $\lesssim5\,\mathrm{Mpc}$, the proposed \textsl{Athena} mission
  holds promise for a detection of the 5.9\,keV line.
Our estimates for the line flux detectabilities (see Fig.~\ref{fig:f2}) 
can be used as a reference for future SN~Ia X-ray observations.
Given the scarcity of SN~Ia events in the local Universe and the sensitivities
of current generation X-ray observatories, it is not possible to constrain 
the relative rates of SNe~Ia from various progenitor models 
\citep[e.g.][]{ruiter2011a} with X-ray line fluxes alone. For individual,
fortuitously nearby SNe, however, the prospects are better. 
In particular, we find that in this case a ${\lesssim} 500\,\mathrm{ks}$
exposure of SNe in the local group (M31) would suffice to detect and
distinguish our two explosion models. 
Observations with lower exposure times (${>} 50$\,ks) should allow for a detection of the 5.9\,keV line at a significance of $5\,\sigma$ in case of the delayed-detonation model.
Thus, observations of the $5.9\,\mathrm{keV}$ line
provide an independent diagnostic tool that can be used
together with measurements at UV, optical, and IR wavelengths
to address the open questions of SN~Ia explosion scenarios and progenitor channels.

\section*{Acknowledgements}
This research has made use of a collection of ISIS scripts provided by the Dr. Karl Remeis observatory, Bamberg, Germany at \url{http://www.sternwarte.uni-erlangen.de/isis/ }. 
We thank Matthias K\"uhnel for his support in calculating the
\textsl{Astro-H}/SXS detector background count rates, Christian
Schmid for his support with the \textsl{Athena/X-IFU} response simulations, and 
J. E. Davis for the development of the \textsl{slxfig} module that has been used to prepare 
some of the figures in this work. The work by KM is partly
supported by the WPI Initiative, MEXT, Japan, and by the Grant-in-Aid for
Scientific Research (23740141, 26800100).The hydrodynamical simulations presented here were carried out in part
at the Forschungszentrum J{\"u}lich with grants PRA042 and HMU13/14
and at the Computer Centre of the Max Planck Society,
Garching. We acknowledge support by the DFG via the Transregional Collaborative Research
Centre TRR 33 ``The Dark Universe'', the Emmy Noether Program \mbox{(RO~3676/1-1),} 
the ARCHES prize of the German Ministry of Education and
Research (BMBF), 
the graduate school ``Theoretical Astrophysics and
Particle Physics'' at the University of W\"urzburg (GRK 1147), the ARC Laureate Grant FL0992131, the
Excellence Cluster EXC~153, the Helmholtz Association
(HGF) through the Nuclear Astrophysics Virtual Institute (VH-VI-417), and 
the European Research Council (ERC-StG EXAGAL-308037). We also thank
the DAAD/Go8 German-Australian exchange programme for travel support.

\end{document}